# Alternating active-dormitive strategy enables overtaking by disadvantaged prey through Parrondo's paradox

Tao Wen[1], Eugene V. Koonin[2], Kang Hao Cheong[1,3*]

*1 Science, Mathematics and Technology Cluster, Singapore University of Technology and Design (SUTD), 8 Somapah Road, S487372, Singapore.*

*2 National Center for Biotechnology Information, National Library of Medicine, National Institutes of Health, Bethesda, MD 20894, USA.*

*3 SUTD–Massachusetts Institute of Technology International Design Centre, Singapore.*

\*Authors for correspondence. E-mail: kanghao_cheong@sutd.edu.sg

**Abstract**

**Background:** Dormancy is a costly adaptive strategy that is widespread among living organisms inhabiting diverse environments. We explore mathematical models of predator-prey systems, in order to assess the impact of prey dormancy on the competition between two types of prey, a perennially active (PA) and capable of entering dormancy (dormitive).

**Results:** Both the active form and the dormant form of the dormitive prey are individually at a disadvantage compared to the PA prey and would go extinct due to their low growth rate, energy waste on the production of dormant prey, and inability of the latter to grow autonomously. However, the dormitive prey can paradoxically outcompete the PA prey with superior traits and even cause its extinction by alternating between the two losing strategies. This outcome recapitulates the game-theoretic Parrondo's paradox, where two losing strategies combine to achieve a winning outcome. We observed higher fitness of the dormitive prey in rich environments because a large predator population in a rich environment cannot be supported by the prey without adopting an evasive strategy, that is, dormancy. In such environments, populations experience large-scale fluctuations, which can be survived by dormitive but not by PA prey.

**Conclusion:** Dormancy of the prey appears to be a natural evolutionary response to self-destructive over-predation that stabilizes evolving predator-prey systems through Parrondo's paradox.

**Background**

Dormancy, such as hibernation in mammals [1, 2], diapause of insects and zooplankton [3], as well as the soil seed banks [4, 5], is a widespread and efficient adaptive strategy for populations under environmental fluctuation. Various types of dormancy are also a common form of stress response in bacteria including sporulation of Gram-positive bacteria [6-8]. The evolutionary stability of dormancy has been investigated under varying environments [3, 9, 10], as well as under the environment with minimal abiotic fluctuations [3]. However, the dormant form of an organism is much more costly compared with the active form in the course of evolution, due to its inability to breed and the large energy consumption during resting egg production as is the case for many invertebrates, such as zooplankton [11, 12].

Hence, dormancy has become a topic of major interest in evolutionary biology: why does dormancy persist and remain a competitive evolutionary strategy? Previous studies have shown the advantage of predator dormancy when two types of predators compete for a single resource [13, 14]. This phenomenon was first experimentally discovered by [15] and subsequently analyzed by [16]. They found that large-amplitude fluctuations can be avoided by predator dormancy in the predator-prey dynamic population model, explaining the paradox of enrichment [17-19]. However, the effect of prey dormancy has not been studied: how will dormancy affect the prey, and can prey dormancy also suppress large-amplitude fluctuation?

In order to identify the factors that determine whether prey remains active or goes into dormancy, it is necessary to quantify the competition among prey. Under high density of predators, the conditions are harsh for prey because they have to avoid being foraged by numerous predators. Under these conditions, the ability to enter dormancy will be beneficial to the prey allowing it to reduce its consumption and breeding, as well as the probability of being discovered. In contrast, the prey will remain active to promote population growth under a low density of predators. Furthermore, dormancy is beneficial for prey (actually, for any organism) under large-amplitude environmental fluctuation, especially under harsh environmental conditions, but not in a safe, stable environment. Recently, it has been shown that environmentally destructive populations can survive by switching between 'nomadic' and 'colonial' forms [20]. Predator dormancy also allows the predator to survive in the large-amplitude fluctuation [14]. In these models, one form grows rapidly but depletes the environment (which includes prey in the case of predator and host in the case of a parasite), whereas the other one does not affect the environment but decays. Each of these strategies individually result in extinction but combined, they can ensure the survival of the population. Therefore, these models exhibit Parrondo's paradox, an abstraction of the phenomenon of flashing Brownian ratchets [21-23], where a winning outcome can be achieved by alternating between two losing strategies [24-26].

Inspired by the previous analyses of the predator-prey model, we propose a population dynamic model to investigate the competition between two forms of prey with different settings under predation. The two forms of dormitive prey $y_1$ and $y_2$ are both losing strategies compared with the perennially active (PA) prey $p$ with superior traits. Indeed, the active form of the dormitive prey $y_1$ has a lower growth rate than the PA prey, spending energy on dormant offspring rather than foraging, and thus loses; the dormant form $y_2$ is also a losing strategy due to its low growth rate and inability to grow on its own. In the game-theoretic perspective, these two strategies cannot individually compete with the PA prey. However, alternating between these two losing strategies allows the dormitive prey to gain advantage in the competition. This result recapitulates Parrondo's paradox in game theory. An additional unexpected finding is that the dormitive prey, which generally would be assumed to be more effective under harsh environmental conditions, in this model, has a higher fitness in richer environments. This outcome is determined by the high density of predators and the large-amplitude fluctuations.

**Results**

**The competition between different forms of prey and the effect of dormancy**

We develop a population dynamic model to explore the competition between the two types of prey under predation (Equations 1 -- 3). The differences between the two types of prey are that (1) the PA prey $p$ has a higher growth rate $r_p$; (2) the dormitive prey has two forms (active form $y_1$ and dormant form $y_2$) which are determined by dormancy switching function $\mu(z)$ and dormancy termination rate $\alpha$. In general, the prey with a higher growth rate will dominate and survive for a long time in the environment under predation. Hence, the density of the predator $z$ is one of the vital parameters for the competition of prey. Especially, the predator density determines the dormancy switching in our model such that there is more dormant form under a higher predator density but more active form under a lower density of predators. In game-theoretic terms, the active form $y_1$ and dormant form $y_2$ of the dormitive prey are both losing strategies compared with the PA prey $p$ because (1) $y_1$ has a lower growth rate and spends extra energy to produce dormant offspring, and (2) $y_2$ also has a lower growth rate and cannot support population growth on its own. However, will the dormitive prey with a lower growth rate but capable of entering dormancy under harsh environmental conditions overtake the PA prey and win the competition in the environment? The detailed descriptions, the initial values, and the units of parameters and functions used in our model, which were obtained from published real-world data [11, 14, 27], are given in Table 1.

In the first competition scenario, $y_1$ or $y_2$ of the dormitive prey compete with $p$ individually (**Figure 1A, 1B**), under different initial population densities, dormancy termination rates $\alpha$, and dormancy switching functions $\mu(z)$. In this case, $y_2$ equals 0 (**Figure 1A**) and $y_1$ equals 0 (**Figure 1B**) individually. In this competition, $y_1$ is disadvantaged compared with $p$ and eventually goes extinct ($t=40$) due to its low growth rate (**Figure 1A**), whereas $y_2$ becomes extinct already in the first wave because it cannot sustain population growth on its own (**Figure 1B**). Thus, both forms of the dormitive prey are losing strategies for $y$, so that $p$ outcompetes $y$ under all conditions. In this competition, the interaction between $p$ and $z$ follows the Rosenzweig-MacArthur criterion [28].

In the next competition scenario, the dormitive prey can switch between $y_1$ and $y_2$ (**Figure 1C**) due to the setting of $\mu(z)$. In this case, $p$ is dominant in the beginning ($t \in [0, 320]$) thanks to its high growth rate $r_p$. However, $p$ cannot support the high density of predators. At high density, the predator $z$ will forage for a large number of $p$, causing its extinction, because $p$ has no recourse to an evasive strategy (dormancy). Thus, $p$ begins to decline after $t = 200$ and becomes extinct at $t = 380$. Then, $y$ overtakes $p$ and gradually evolves towards a steady state of coexistence with $z$. There are more active forms $y_1$ at lower density of $z$ and more dormant forms $y_2$ under higher density of $z$ because of the setting of $\mu(z)$. The density of $y_1$ is similar to the density of $p$, which is determined by the carrying capacity $K$. In order to examine the dynamics of different populations more closely, a cycle of the dynamic process ($t \in [400, 450]$) is enlarged (**Figure 1D**). The peaks of different populations do not occur at the same time. Specifically, the peak time of $y_2$ occurs between the peak times of $y_1$ and $z$. In addition, the growth rate of $y_2$ around $t = 420$ is higher than the rate during $t \in [410, 418]$, which is determined by the growth term $r_y \left(1 - \frac{p + y_1}{K}\right)(1 - \mu(z)) y_1$. During $t \in [410, 418]$, the growth of $y_2$ is mainly due to the increasing density of $y_1$, causing the increase of $\left(1 - \frac{p + y_1}{K}\right) y_1$, under the logistic growth model. However, after $t = 418$, the density of $y_1$ begins to drop and the density of $z$ begins to increase, and therefore, the growth rate of $y_2$ is higher than before because of the rapid increase in $1 - \mu(z)$ (Equation 3). After $t = 420$, the growth rate is slower than the foraging rate, so $y_2$ declines and $z$ reaches its peak.

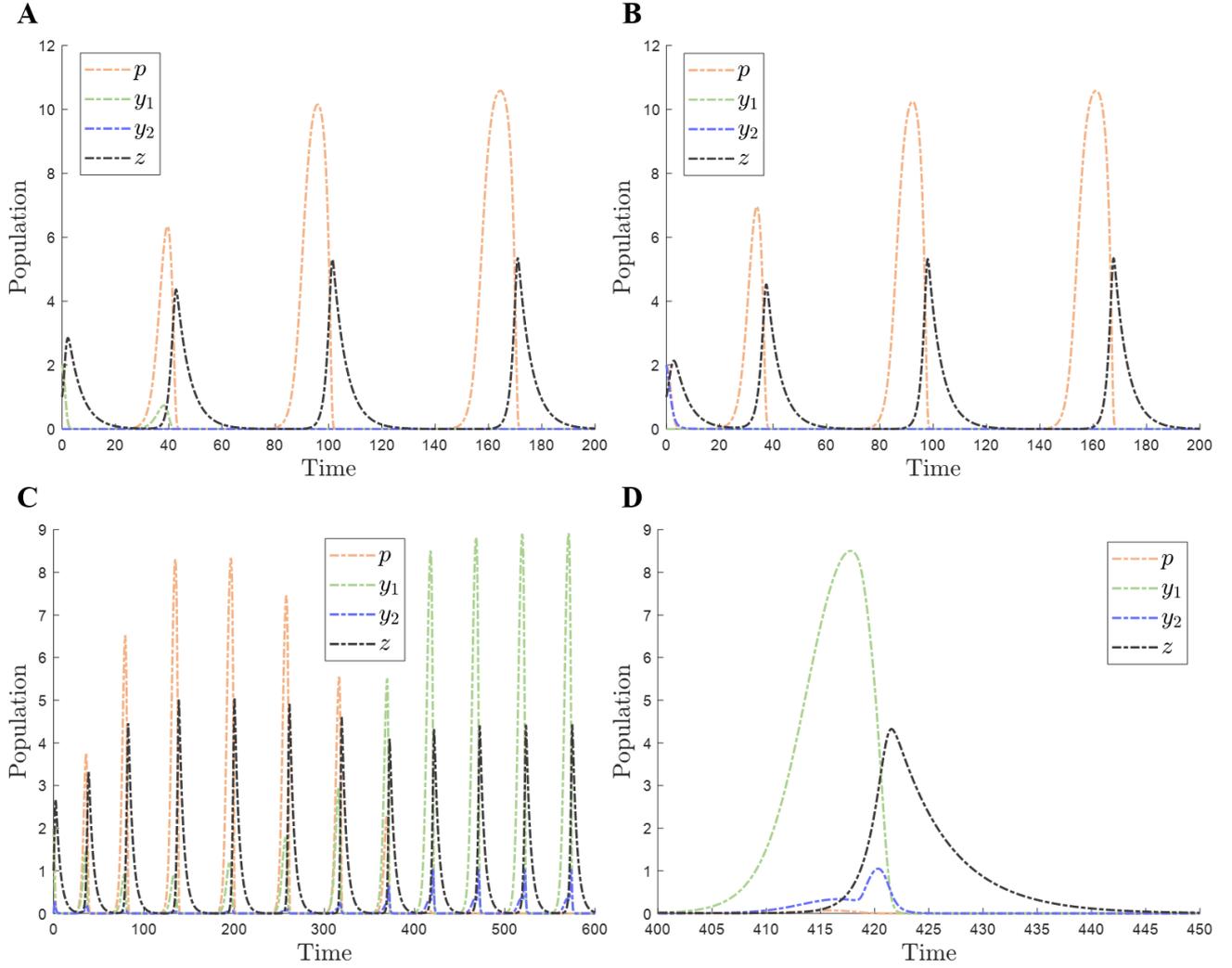

**Figure 1 | Competition between two types of prey.**

The parameter values are from Table 1 unless stated otherwise.

a, Only the $y_1$ form of the dormitive prey is present, in competition with $p$. The initial population density is $[2,2,0,1]$, $\alpha=1$, and $\mu(z)=1$.

b. Only the $y_2$ form of the dormitive prey is present, in competition with $p$. The initial population density is $[2,0,2,1]$, $\alpha=0$, and $\mu(z)=0$.

c. Both $y_1$ and $y_2$ forms of the dormitive prey are present, in competition with $p$. The initial population density is $[2,2,0,1]$.

d, A segment of the dynamics in Figure 1C is enlarged to show the evolution of each population ($t \in [400, 450]$).

**Impact of the carrying capacity on the outcome of the competition**

As the carrying capacity $K$ increases, there will be more prey, causing a higher density of predators. However, our previous results show that $p$ cannot support the high density of $z$ and thus becomes extinct. Hence, the impact of the carrying capacity on the competition between the two types of prey should be studied. Compared with $K=11$ (**Figure 1C**), the dormitive prey will take advantage earlier at $K=13$ (**Figure 2A**). $p$ cannot reach its peak (near carrying capacity) under these conditions and will begin to decline at a point where $p$ still has an apparent potential to rise. This is the case because the current density of prey already results in high predator density such that the prey cannot reach the environmental capacity due to the high rate of predation. However, $z$ grows faster than before thanks to the availability of adequate resources, so $p$ will become extinct quickly because of the lack of an evasive strategy. The first time when the peak density of $y_1$ is higher than the peak density of $p$ is called the reversal time $T_w$. The reversal time $T_w$ is 195.1 in this case (**Figure 2A**), that is, reversal occurs earlier than it does at $K=11$. When $K$ continues to rise to 20 (**Figure 2B**), the reversal time further shortens, to $T_w=89.2$. In addition, the peak density of $y_1$ is also higher than for the previous conditions and comes close to the carrying capacity. In contrast, when $K$ decreases to 9 (**Figure 2C**), $p$ is at an advantage because the density of $z$ is lower than it is at larger $K$. This allows $p$ to grow in the environment and avoid extinction. The peak density of $p$ is also close to $K=9$, the same as discussed before. At a high predator density, the PA prey will be extensively foraged and will eventually go extinct. In contrast, the dormitive prey can enter dormancy to survive at a high density of $z$, which is a competitive survival strategy. Hence, the dormant form is advantageous to the prey under population fluctuation in the apparent competition [29]. The reversal time $T_w$ changes rapidly when the carrying capacity $K$ is low but stays nearly constant at high $K$ (**Figure 2D**).

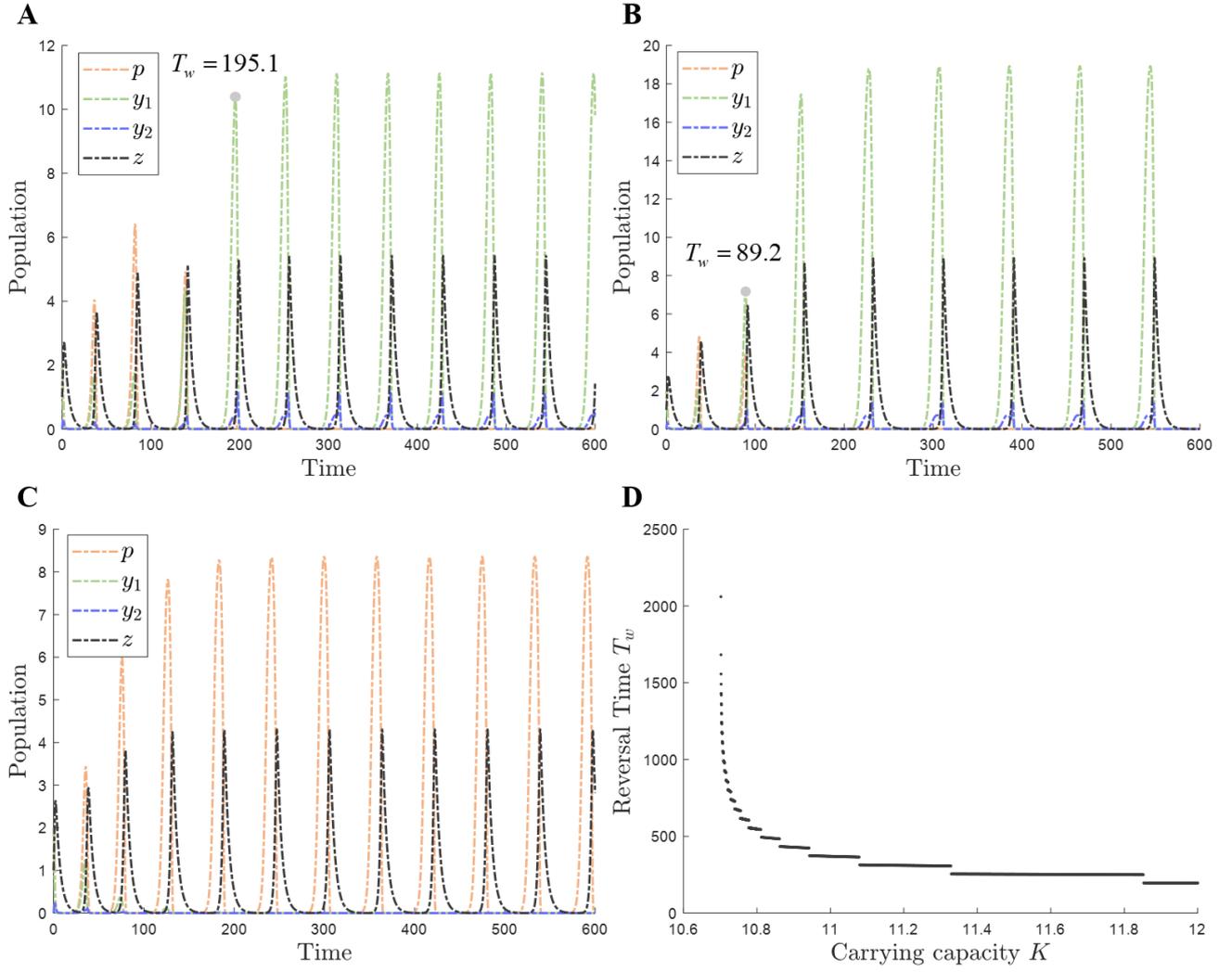

**Figure 2 | Competition between two types of prey under different carrying capacities, and the reversal time $T_w$.**

The parameter values are from Table 1 unless stated otherwise. The initial population density is $[2,2,0,1]$.

a, Carrying capacity $K$ equals $13$.

b, Carrying capacity $K$ equals $20$, causing higher density of each population and shorter reversal time $T_w$ compared to the case that $K$ equals $13$.

c, Carrying capacity $K$ equals $9$, $p$ wins the competition.

d, The relationship between the reversal time $T_w$ and carrying capacity $K$.

### Analysis of the parameter space

We further explored the parameter space of the model to determine how the dormitive prey gains advantage under different conditions. The outcomes were evaluated by the highest density of each competitor during $t \in [2800, 3000]$ (stable competition). Due to the impracticality of all-vs-all parameter comparison, the parameters were analyzed in pairs as follows.

The carrying capacity $K$ was first compared with the dormancy termination rate $\alpha$ (**Figure 3A**). When $K$ is low, $p$ reaches higher abundance than $y_1$, no matter how $\alpha$ changes. This is because there is not enough of $z$ to consume resources (prey) under low $K$, thus avoiding the extinction of $p$. However, as $K$ increases, a large number of $z$ will forage for $p$, leading to the extinction of $p$. In this case, $y_1$ gains advantage under medium values of $\alpha$, because both the high and low $\alpha$ only allow a single form of dormitive prey to exist, which are losing strategies compared with the PA prey (**Figure 1A**, **B**). The carrying capacity $K$ is then compared with the predator death rate $d_z$ (**Figure 3B**). With the increase of $K$, the density of both types of prey will increase, the same as the result in **Figure 3A**. At low $K$, only small values of $d_z$ allow $y_1$ to win because only the dormitive prey can survive under the high density of predators in this case. There will be more prey in the environment with higher $K$, leading to many predators, thus, $p$ can only win under higher values of $d_z$ (fewer predators) as shown above. Hence, there is only a narrow range of $d_z$ for $p$ to dominate in the environment under high $K$, but a wide range of $d_z$ under low $K$.

Four parameters in the dormancy switching function $\mu(z)$ were then compared because of the importance of switching for $y$. The lower bound $\chi$ and range $\varphi$ determine how many prey remain active and how many go dormant (**Figure 3C**). The values of $\chi$ and $\varphi$ obey the condition $0 \leq \chi \leq \chi + \varphi \leq 1$, which means that the value of the dormancy switching function should be between 0 and 1. We find that $y_1$ can only gain advantage when $\chi$ is relatively low ($\chi \leq 0.6$) and $\chi + \varphi$ is close to 1. $\chi$ cannot be too large because the prey need a certain amount of active form to provide the growth capacity in all cases, otherwise $y$ will go extinct. The reason why $\chi+\varphi\approx 1$ is to ensure full reproduction under safe conditions, so that the dormitive prey can survive under environmental fluctuations. In the comparison of switching threshold $\eta$ and width $\sigma$ in $\mu(z)$ (**Figure 3D**), the dormitive prey can reach a higher abundance than $p$ with low $\sigma$ because of the sharp switching. The sharp switching between the active and dormant form results in a strong fitness to the environment, allowing the disadvantaged prey $y$ to win in the competition.

The relationships between $z$ and two types of prey are further studied. The comparison between the predator growth efficiencies from the active form $k_p, k_{y_1}$ and the efficiency from the dormant form $k_{y_2}$ (**Figure 3E**) shows that $k_{y_2}$ does not affect the result of the competition. Higher $k_p$ and $k_{y_1}$ will cause more $z$ in the environment, resulting in the winning outcome of $y_1$ (similar results in **Figure 3B**). We assume the foraging efficiencies $c_{y_2} \leq c_p$ and $c_{y_2} \leq c_{y_1}$ in **Figure 3F** because the dormant form is less likely to be found and thus avoids predation. There will be more $z$ under higher $c_p$ and $c_{y_1}$, and thus, it can be concluded that $y$ will win in this case (same reason as in **Figure 3B**).

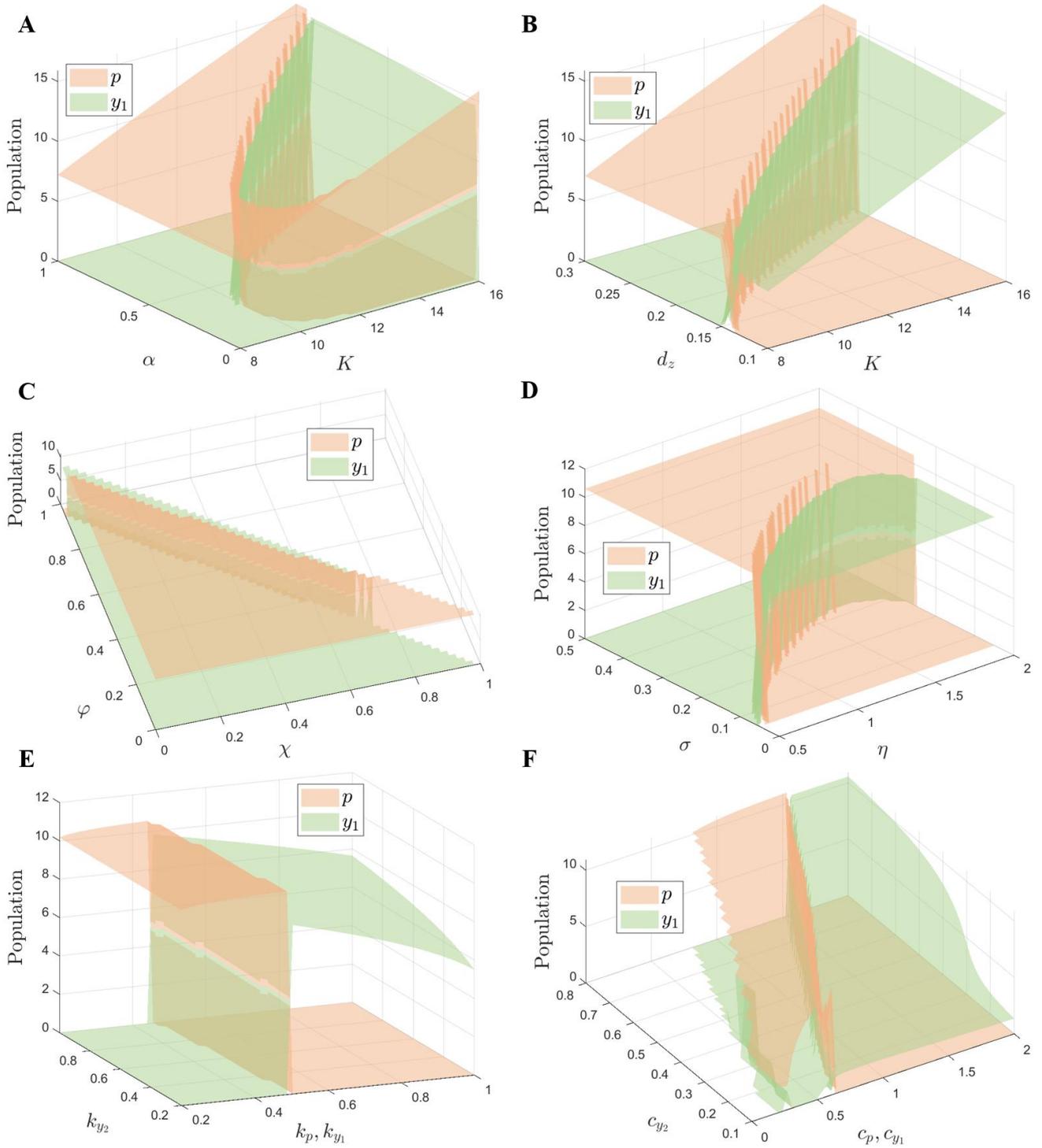

**Figure 3 | The parameter space that results in Parrondo's paradox in the competition.**

The parameter values are from Table 1 unless stated otherwise.

a-b, Joint impact of carrying capacity $K$, dormancy termination rate $\alpha$, and predator death rate $d_z$ on prey density.

c-d, Joint impact of parameters in the dormancy switching function $\mu(z)$ on prey density, including the lower bound $\chi$, range $\varphi$, switching threshold $\eta$, and width $\sigma$.

e-f, The relationships between the predator and two types of prey, including the growth efficiencies $k_q$ and foraging efficiencies $c_q$, are compared.

**Discussion**

Analysis of the present model shows that the dormitive prey can outcompete the PA prey with superior traits by switching between the active and dormant form. The counterintuitive results obtained here are that (1) the dormitive prey with lower growth rate and extra energy loss can overtake the PA prey and win the competition; (2) prey dormancy is advantageous in rich environments (high carrying capacity), whereas intuitively it could be expected to gain advantage in poor environments. These surprising outcomes of the competition can be explained through Parrondo's paradox: alternation of two losing strategies can result in a winning strategy. In this model, both the active form and the dormant form of the dormitive prey are individually disadvantaged in the competition against the PA prey because (1) the active form of the dormitive prey has a lower growth rate and spends extra energy to produce dormant offspring; (2) the dormant form also has a lower growth rate and cannot support population growth on its own (**Figure 1A, 1B**). However, a winning outcome for the dormitive prey can be achieved by alternating between the active form and the dormant form (**Figure 1C**). This result can be explained by Parrondo's paradox. Parrondo's paradox also might be applicable to other biological competitions. For example, bacteriophages can switch between active reproduction (host-destroying lytic phase) and dormancy (lysogenic phase) depending on the state of the external environment [30], and the dormancy of phages (lysogeny) can be considered an adaptation to host population oscillations.

Under a high carrying capacity, a high density of prey is reached by logistic growth [31], thereby supporting a high density of predators. However, the high density of predators will lead to the extinction of the PA prey because the latter has no recourse to dormancy as an evasive strategy (**Figure 2A, 2B**). Hence, the dormitive prey has higher fitness under higher carrying capacity (higher density of predators). Furthermore, the higher the carrying capacity, the sooner (smaller value of the reversal time $T_w$) the dormitive prey will gain advantage and overtake (**Figure 2D**) because predators will grow faster and forage more PA prey. In contrast, the PA prey will gain advantage under low peak density of predators (**Figure 2C**) because it can support the predator population at this low density. Therefore, the density of predators (determined by the carrying capacity in our model) is the key factor that determines the outcome of the competition between the two types of prey [29, 32]. This allows us to interpret the second counterintuitive result: prey dormancy is advantageous in richer environments. The resource-rich environment not only increases the density of the prey, but also causes the large-amplitude population (predator-prey) cycles and high density of predators due to the prey oversupply, followed by over-predation [15, 17]. The dormant form of prey can survive under the high density of predators in the population cycles to come back subsequently, whereas the PA prey goes extinct in the cycles. Furthermore, no matter which prey has advantage in the competition, the peak density of prey is higher in richer environment, approaching the carrying capacity and causing large amplitude cycles, which is beneficial to the dormitive prey.

The large-amplitude population cycles have been explored in diverse real-world cases, such as the relationship between Daphnia and plankton [15, 33], Arctic lemmings and weasels [34], parasites and mountain hares [35], as well as intraspecific competition of Antarctic krill [36]. These results show the effect of over-predation and potential applications of Parrondo's paradox [37, 38]. The large-amplitude fluctuations are usually accompanied by extinction of populations [16, 17], but dormancy allows the prey to survive the fluctuation, with a subsequent comeback. Thus, our key finding is that the dormitive prey can gain advantage and overtake in the competition under large-amplitude fluctuations. Indeed, prey dormancy has been observed in a variety of predator-prey systems including bacteria (Myxococcus as predator and Bacillus as prey) [39-41], arthropods—spider mites [42, 43], and small rodents—dormice [44, 45].

An obvious limitation of this work is that, because of the wide variation of the parameters of predator and prey populations, the evolutionary dynamics derived from the real-world data [11, 14, 27] might substantially differ from that predicted by the model. Nevertheless, we explored in detail the impact of multiple parameter combinations on the competition outcome. In particular, this analysis shows that the parameter values that cause a higher density of predators, such as lower death rate, higher predator growth efficiencies, and higher foraging efficiencies, will lead to the extinction of the PA prey. In addition, the range of dormancy termination rate ensuring the winning outcome for the dormitive prey is determined by the carrying capacity, which can also affect the density of predators. The parameters in the threshold-induced dormancy function also need to be in a suitable range for the dormitive prey to win the competition.

**Conclusion**

The analysis of the present model indicates that the counterintuitive winning outcome for the dormitive prey is underlain by the game-theoretic Parrondo's paradox. In the large-amplitude fluctuation, dormancy can help the dormitive prey to outcompete the PA prey with superior traits and even cause the extinction of the PA prey. Parrondo's paradox is likely to be widely applicable to other biological competitions with large-amplitude fluctuations.

**Methods**

**Population Model**

In the dynamic population model, we introduce an additional prey population ($y$) to the existing predator-prey model, adapted from the Lotka–Volterra-derived model of Rosenzweig and MacArthur [28, 46, 47]. In detail, two prey populations ($p, y$) are both preyed on by the predator population ($z$). $p$ is the density of PA prey, and $y = y_1 + y_2$ is the density of dormitive prey, with $y_1$ and $y_2$ corresponding to the active and dormant subpopulations, respectively. The differential equations of

this proposed model are

$$\dot{p} = r_p\left(1 - \frac{p + y_1}{K}\right)p - f_p(p)z,$$

$$\dot{y}_1 = r_y\left(1 - \frac{p + y_1}{K}\right)\mu(z)y_1 + \alpha y_2 - f_{y_1}(y_1)z, \quad (1)$$

$$\dot{y}_2 = r_y\left(1 - \frac{p + y_1}{K}\right)(1 - \mu(z))y_1 - \alpha y_2 - f_{y_2}(y_2)z,$$

$$\dot{z} = k_p f_p(p)z + k_{y_1} f_{y_1}(y_1)z + k_{y_2} f_{y_2}(y_2)z - d_z z.$$

Two types of prey follow the logistic growth model [48] with the environment carrying capacity $K$ and different maximum growth rates ($r_p$, $r_y$), as well as the predation from $z$ at rates $f_p(p)$, $f_{y_1}(y_1)$, and $f_{y_2}(y_2)$ (Equation 2). The active form $y_1$ is also supplemented by the dormant form $y_2$ at the termination rate $\alpha$, and only a fraction of energy, $\mu(z)$, is used for the active descendant of the dormitive prey. Therefore, the remaining fraction, $1 - \mu(z)$, is used for the dormant progeny. The predator $z$ grows in proportion to the growth efficiencies ($k_p$, $k_{y_1}$, $k_{y_2}$) and predation rates ($f_p(p)$, $f_y(y_1)$, $f_{y_2}(y_2)$), while dying at rate $d_z$. Recent published real-world data [11, 14, 27] shown in Table 1 are used to validate our proposed dynamic model. These data have been applied to describe the traits of prey and predator in the competition from the plankton modelling literature. Numerical simulations were implemented by explicit Runge-Kutta (2,3) methods. The accuracy of the simulation is ensured by the strict tolerance level in the repeated experiment, which makes the final coefficient not change significantly and less than 1%. In the simulation, the relative error tolerance and absolute error tolerance are both $10^{-8}$.

**Parameter Functions**

The predation function $f_q(q)$ originated from the Holling type II function response, a monotone increasing function based on the property of prey,

$$f_q(q) = \frac{c_q q}{1 + c_q h_q q}, q \in \{p, y_1, y_2\}, \tag{2}$$

where $c_q$ and $h_q$ are the foraging efficiency and handling time, respectively. There are different values of parameters for different types of prey. $f_q(q)$ equals to $c_q q$ and $1/h_q$ when $q \to 0$ and $q \to \infty$, respectively.

The dormancy switching function $\mu(z)$ is obtained by the improved sigmoid switching function (monotonically decreasing) based on the predator density $z$,

$$\mu(z) = \chi + \varphi \left[1 + \exp\left(\frac{z - \eta}{\sigma}\right)\right]^{-1}, \tag{3}$$

where $\chi$ and $\varphi$ are the lower bound and range of the function, $\eta$ and $\sigma$ denote the switching threshold and shape of the switching function, respectively.

**Table 1 | The descriptions, values, and unite of parameters and functions used in the model (from recent real-world data [11, 14, 27]).**

| Parameter | Description | Value | Units |
|---|---|---|---|
| $p$ | Perennially active prey density | 2 | $mgL^{-1}$ |
| $y_1$ | Dormitive prey (active form) density | 2 | $mgL^{-1}$ |
| $y_2$ | Dormitive prey (dormant form) density | 0 | $mgL^{-1}$ |
| $z$ | Predator density | 1 | $mgL^{-1}$ |
| $r_p, r_y$ | Prey growth rate | 0.55, 0.5 | $mgL^{-1}$ |
| $K$ | Prey carrying capacity | 11 (Variable) | $mgL^{-1}$ |

| Symbol | Description | Value | Units |
|---|---|---|---|
| $d_z$ | Predator death rate | 0.2 | $Day^{-1}$ |
| $\alpha$ | Dormancy termination rate | 0.05 | $Day^{-1}$ |
| $k_p, k_{y_1}$ | Predator growth efficiencies (from active form) | 0.5 | --- |
| $k_{y_2}$ | Predator growth efficiency (from dormant form) | 0.25 | --- |
| $c_p, c_{y_1}, c_{y_2}$ | Predator foraging efficiencies | 1, 1, 0.4 | $Day^{-1}mg^{-1}L$ |
| $h_p, h_{y_1}, h_{y_2}$ | Predator handling times | 0.5 | $Day$ |
| $\eta$ | Dormancy switching threshold | 1 | $mgL^{-1}$ |
| $\sigma$ | Dormancy switching width | 0.1 | $mgL^{-1}$ |
| $\chi$ | Lower bound | 0.2 | $mgL^{-1}$ |
| $\varphi$ | Range | 0.75 | $mgL^{-1}$ |
| $f_p(p), f_{y_1}(y_1), f_{y_2}(y_2)$ | Predation rates | Function | $Day^{-1}$ |
| $\mu(z)$ | Dormancy switching function | Function | --- |

## Abbreviations

PA: perennially active.

## Acknowledgements

Not applicable.

## Funding

T. W. and K. H. C. are supported by the Singapore University of Technology and Design (Grant

No. SRG SCI 2019 142). E.V.K. is supported by the Intramural Research Program of the National Institutes of Health of the USA (National Library of Medicine).

## Author Contributions

K. H. C. and E.V.K. designed research; T. W. and K. H. C. performed research; T. W., K. H. C., and E.V.K. analyzed the data. T. W., K. H. C., and E.V.K. wrote the manuscript. K.H.C. supervised the study.

## Ethics approval and consent to participate

Not applicable.

## Consent for publication

Not applicable.

## Competing interests

The authors declare no conflict of interest.